\documentclass[aps,prb,reprint,showpacs,superscriptaddress]{revtex4-1}

\usepackage{graphicx}
\usepackage{bm}
\usepackage{dcolumn}
\usepackage{amssymb}
\usepackage{amsmath}
\usepackage{amsfonts}
\usepackage{color}
\usepackage{epsfig}
\usepackage{hyperref}
\usepackage{verbatim}

\def \da{^{\dagger}}

\def \g{\mathrm{g}}

\makeatletter

\newenvironment{figurehere}
  {\def\@captype{figure}}
  {}
\makeatother

\begin{document}

\title{ Suppression of dephasing and phase lapses in the fractional quantum 
Hall regime }

\author{Yehuda Dinaii}
\affiliation{Department of Condensed Matter Physics, The Weizmann Institute of
             Science, Rehovot 76100, Israel}
\author{Yuval Gefen}
\affiliation{Department of Condensed Matter Physics, The Weizmann Institute of
             Science,  Rehovot 76100, Israel}
\author{Bernd Rosenow}
\affiliation{Physics Department, Harvard University, Cambridge, Massachusetts 02138, USA}
\affiliation{Institut f\"ur Theoretische Physik, Universit\"at Leipzig, D-04103,
             Leipzig, Germany}

\begin{abstract}

A charge fluctuator which is electrostatically  coupled to a conducting channel may fully dephase quantum transport through the latter. Here, we address the case where a quantum dot (QD), playing the role
of a charge  fluctuator, is tunnel-coupled to an additional channel. In the 
case where the latter may  support fractional charge, distinct differences 
from the integer case arise: Abrupt phase lapses of
the transmission through the conducting channel occur (which may or may not be equal to $\pi$). This is accompanied by a cusp-like suppression of the interferometer's visibility, yet no full dephasing.
We interpret our findings in terms of the entanglement between the fluctuator 
and the conducting channels.

\end{abstract}


\maketitle

Interference between possible paths is a fundamental quantum mechanical 
phenomenon. Its realization via an electronic Mach-Zehnder interferometer 
(MZI)~\cite{Ji_2003, *Neder_2006, *Roulleau_2007, *Litvin_2007, *Litvin_2008, 
*Roulleau_2008, *Bieri_2009, *Roulleau_2009} and an electronic Fabry-P\'erot 
interferometer~\cite{Camino_2007, *Zhang_2009, *Ofek_2010, *McClure_2012} 
plays an important role in investigating basic physics phenomena such as the 
Aharonov-Bohm effect, action-free measurement, correlations among particles, 
and fractional statistics of Abelian and non-Abelian anyons.  Understanding 
the degradation of quantum interference (dephasing) is crucial for 
nanoelectronic technologies, and is of profound importance for clarifying  the 
elusive transition from a quantum behavior to a classical one. Dephasing of 
interferometry signal due to the interaction with a detector is intimately 
related to the entanglement between them. It is often associated with 
acquiring information about which path the interfering particle has 
taken.~\cite{Neder_2007_PRL, *Neder_2007_NaturePhysics, Levinson_1997, 
Aleiner_1997, Gurvitz_1997}

It is common practice to model the effect  of an environment on a system 
through  a bath of harmonic oscillators or a puddle of Fermi liquid coupled to 
the latter. However, recent developments in experimental techniques have made 
it possible to perform controlled-dephasing experiments, where the dephasor 
consists of only few degrees of freedom, and can be controlled with high 
precision. This has introduced the need for models of dephasors that are 
non-Gaussian and fully quantum mechanical. An out-of-equilibrium detector 
capable of determining the path taken by the electron, and thereby destroying 
the interference signal, is a standard paradigm in mesoscopic physics.  
Recently it has been shown that the Friedel sum rule is a useful tool for 
analyzing dephasing induced by charge fluctuations~\cite{Rosenow_2012}.

\begin{figure}[tbp]
\begin{center}
\vspace{0cm}
\includegraphics[width=0.45\textwidth]{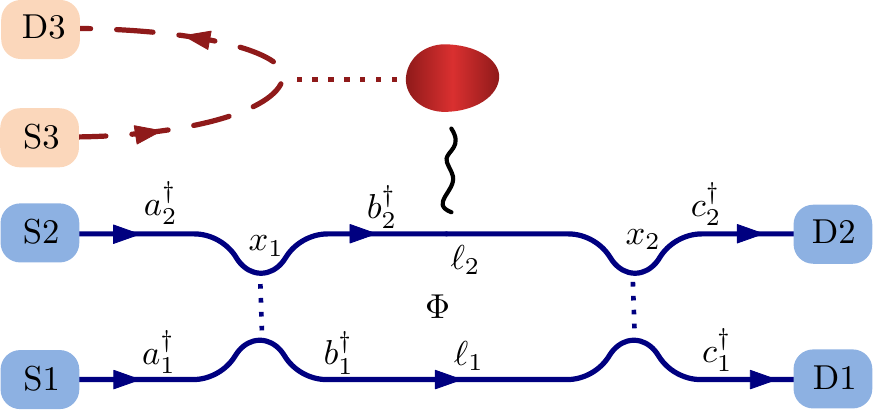}
\vspace{0cm}
\caption[] { \label{fig:setup}
Schematics of the setup. A quantum dot (QD) (red solid puddle) is capacitively 
coupled (black wiggly line) to an arm of a Mach-Zehnder interferometer (MZI) 
(blue solid lines). The QD is tunnel-coupled (red dots) to a chiral Luttinger 
liquid lead (red dashed line). Tunneling between the two MZI arms is denoted 
by blue dotted lines. Magnetic flux $\Phi$ is enclosed by the ``lower'' and 
``upper'' MZI arms whose lengths are  $\ell_1$ and $\ell_2$, respectively. The 
arrows denote the direction of the chiral motion from the sources (``S'') 
towards the drains (``D'').  Creation operators associated with the various 
regions of each channel are designated in the figure.
\vspace{-0.5cm}
}
\end{center}
\end{figure}

Here we study a paradigmatic controlled-dephasing setup of a MZI interacting 
with a fluctuator, but with a twist: the fluctuator operates in the fractional 
quantum Hall regime. The fluctuator is realized by a quantum dot (QD) (cf.  
Fig.~\ref{fig:setup}), which is electrostatically coupled to the upper arm of 
the MZI. Since the QD has two charging states (``empty'' and ``occupied''), 
the fluctuation pattern of the charge on it defines quantum telegraph noise.  
Consequently, the phase accumulated along the upper arm of the MZI fluctuates 
between two values ($\theta$ and $\theta + \delta$).  The QD is weakly 
tunnel-coupled to a channel that supports fractional charge.  This is to be 
contrasted with previous studies~\cite{Grishin_2005,
*Yurkevich_2010, Neder_2007_NewJournalOfPhysics, *Abel_2008, Rosenow_2012, 
Weisz_2012}, which examined a QD that was tunnel-coupled to regular electron 
reservoirs. We find that: (i) When the gate voltage applied on the QD is 
varied smoothly, a sharp jump in the phase of the transmission amplitude 
through the upper arm of the MZI occurs (cf.  
Fig.~\ref{fig:vis_phase_occupation}). (ii) For a symmetric arrangement of the 
QD ($\delta = \pi$, see below), the abrupt phase lapse is $\pi$.  (iii) Unlike 
the integer case, even for $\delta \ne \pi$ (asymmetric arrangement), a 
\emph{sharp} phase lapse will appear, which differs from $\pi$. (iv) The 
visibility of the MZI is minimal at the point where a phase lapse occurs, but 
generically it cannot be fully suppressed. (v)  The visibility is a 
monotonically decreasing function of the entanglement between the MZI and the 
QD (cf.  Fig.~\ref{fig:visbility_AAFO_concurrence}).  The protection against 
full dephasing is a consequence of the inability of the MZI and the QD to 
fully entangle.

Consider a quantum Hall state supporting two co-propagating edge modes, for 
instance, filling factor $\nu = 4/3$. These two edge modes can be utilized for 
constructing a MZI, capacitively coupled to a nearby QD~\cite{Weisz_2012}.  
Fig.~\ref{fig:setup} shows schematically such a setup. The MZI consists of two 
interfering chiral edges --- associated with the edge modes of the first fully 
occupied Landau level --- enclosing an Aharonov-Bohm flux, $\Phi$. The QD is 
tunnel-coupled to another channel associated with the edge modes of the second 
Landau level. In our analysis, the QD is modeled by a single electronic level, 
and the tunnel-coupled lead supports fractional charge. Intra- and inter-edge 
interactions within the MZI are neglected.

The interaction between an electron passing through the MZI and the charge on 
the QD is important. Its effect is to provide the partial wave passing through 
the upper MZI arm with an extra phase shift $\delta$ when the QD is occupied. 
This extra phase is a consequence of the variation of the Fermi sea at the 
upper arm of the MZI due to screening of the QD's potential.  Specifically, it 
was shown~\cite{Rosenow_2012} that $\delta / 2\pi$ is the fraction of electron 
charge screened by the upper MZI arm when the QD is occupied by one electron.  
The rest of the electron's charge is screened by other metallic gates/leads, 
which are not part of our model. Thus, fluctuations of the charge residing on 
the QD are accompanied by fluctuations in the transmission phase, which act 
towards dephasing the interference signal of the MZI. 

To evaluate the visibility and the transmission phase through the upper MZI 
arm, which are experimentally measurable quantities, we solve the scattering 
problem for an electron injected at one of the MZI sources.  To this end we 
define creation operators associated with  each of the three regions of the 
MZI channels (cf.  Fig.~\ref{fig:setup}).  These are $a^{\dagger}$, 
$b^{\dagger}$ and $c^{\dagger}$ for $x < x_1$, $x_1 < x < x_2$ and $x > x_2$, 
respectively.  Operators associated with the lower (upper) arm are denoted by 
the subscript 1 (2). Scattering at the two junctions of the MZI enable 
tunneling between the channels.  The relation between the creation operators 
in the three regions is given by $(b_{1}^{\dagger}\; b_{2}^{\dagger}) = 
(a_{1}^{\dagger}\; a_{2}^{\dagger}) s_{1}^{\dagger}$ and $(c_{1}^{\dagger}\; 
c_{2}^{\dagger}) = (b'{}_{1}^{\dagger} \; b'{}_{2}^{\dagger} ) 
s_{2}^{\dagger}$,  where the scattering matrices are $\ensuremath{s_{i} = 
\left( \begin{smallmatrix} r_{i} & t_{i}'\\ t_{i} & r_{i}' 
\end{smallmatrix}\right)}$
with $i = 1,2$. Primes in the second relation denote that additional phases 
are accumulated between the two junctions (see below). Note that the operator 
$b_{2}^{\dagger}$ creates a state before the possible interaction with the QD. 

In reality, the initial state of the subsystem composed of the QD and its 
tunnel-coupled lead may be a very complicated many-body wave function, since 
the second Landau level is partially filled.  However, the MZI interacts 
electrostatically only with the QD, and not with the electrons residing in the 
tunnel-coupled lead. Therefore, the only relevant quantity is the part of the 
wave function that is localized within the QD.  This allows us to employ an 
effective description, whereby the wave function of the QD is that of an 
isolated two-state system, i.e. a qubit. Any state of this qubit can be 
written as $ \left|{\mathrm{QD}}\right\rangle = e^{i \alpha_{1}} \sqrt{n} 
\left| 1 \right\rangle + e^{i \alpha_{2}} \sqrt{1-n} \left| 0 \right\rangle$.  
Here $\left| 1 \right\rangle = d\da \left| 0 \right\rangle$, where $d\da$ 
denotes the creation operator of the QD's electronic state, and $\alpha_1$, 
$\alpha_2$ are phases.  The probability of the QD to be occupied, $n$, equals 
the average occupation of the QD.  Formally,  $n = \left\langle \mathrm{GS} 
\right| d^{\dagger} d \left| \mathrm{GS} \right\rangle$, where $\left| 
\mathrm{GS} \right\rangle$ denotes the ground state of the QD and its 
tunnel-coupled lead.  The QD's level has energy $\varepsilon$, which is 
measured relative to the Fermi energy and can be controlled by a gate voltage.  
In what follows, we treat the QD+lead subsystem as a non-dynamical 
environment~\cite{Stern_1990}, namely the occupation of the QD is unchanged 
during the passage of an electron through the upper MZI arm.  This assumption 
is applicable when the time-of-flight through the MZI is much shorter than the 
typical time of charge fluctuations in the QD. This regime is realized, e.g.  
in Ref.~[\onlinecite{Weisz_2012}]. In addition, we use the inert band 
approximation and do not consider the renormalization of the tunnel coupling 
between QD and Luttinger liquid due to the Coulomb interaction with the MZI.

We now turn to solving the scattering problem. Suppose that current is 
injected at S2 into the upper arm of the MZI.  The wave function of the 
composite system for $x < x_1$ is
\begin{equation}
  \left| \chi_a \right\rangle = \left(e^{i\alpha_{1}} \sqrt{n} \, d^{\dagger} 
  + e^{i\alpha_{2}} \sqrt{1-n} \right) a_{2}^{\dagger} \left| \g \right\rangle 
  \,,
  \label{initial_state}
\end{equation}
where $\left| \g \right\rangle$ is the ground state of the MZI+qubit 
composite. Using the scattering matrix $s_1$, one can express 
$a_{2}^{\dagger}$ in terms of $b_{1}^{\dagger}$ and $b_{2}^{\dagger}$, and 
thus write the wave function for $x = x_1 + 0$.

The phase accumulated along the lower (upper) arm due to the optical path 
$\ell_1$ ($\ell_2$) is denoted by $\phi_{\mathrm{d(u)}}$. The Aharonov-Bohm 
phase $\phi_{\mathrm{AB}} = 2 \pi \Phi / \Phi_0$, where $\Phi_0$ is the 
magnetic flux quantum, is ascribed to the lower MZI arm. We also recall that, 
if the QD is occupied, the extra phase accumulated along the upper MZI arm is 
$\delta$.  Thus, the wave function of the entire system corresponding to an 
electron at $x = x_2 - 0$ is
\begin{eqnarray}
  \left| \chi_b \right\rangle =  e^{i(\alpha_{1} + \phi_{\mathrm{d}} + 
  \phi_{\mathrm{AB}})} \left\{ \sqrt{n} \left[t_{1}' d^{\dagger} 
    b'{}_{1}^{\dagger} + r_{1}' e^{i(\phi_{\mathrm{u}} - \phi_{\mathrm{d}} - 
    \phi_{\mathrm{AB}} + \delta)} d^{\dagger} b'{}_{2}^{\dagger} \right] 
    \right.  \nonumber \\
  \left.  +  \sqrt{1-n} e^{i(\alpha_{2} - \alpha_{1})}\left[t_{1}' 
    b'{}_{1}^{\dagger} + r_{1}' e^{i(\phi_{\mathrm{u}} - \phi_{\mathrm{d}} - 
    \phi_{\mathrm{AB}})} b'{}_{2}^{\dagger} \right] \right\} \Big| \g 
    \Big\rangle \,. \;\;\; \quad
    \label{eq:state_in_front_of_QPC2}
\end{eqnarray}

Employing the scattering matrix $s_2$  one finally obtains the wave function 
of the entire system with $x = x_2 + 0$,
\begin{subequations}\label{eq:WF_after_QPC2}
  \begin{align}
  \left| \chi_c \right\rangle & = e^{i(\alpha_{1} + \phi_{\mathrm{d}} + 
  \phi_{\mathrm{AB}})} \left( \hat{A}c_{1}^{\dagger} + 
  \hat{B}c_{2}^{\dagger}\right) \Big| \g \Big\rangle \,,
  \label{eq:WF_after_QPC2_main}\\
  \hat{A} & = \left[-\sqrt{n}d^{\dagger} + \sqrt{1-n 
  }e^{i(\alpha_{2}-\alpha_{1})}\right]t_{1}'r_{2}\nonumber \\
  & +e^{i \alpha_3 } \left[ -\sqrt{n} e^{i\delta} d^{\dagger} + \sqrt{1-n} 
    e^{i(\alpha_{2} - \alpha_{1})} \right] r_{1}' t_{2}'\,,
  \label{eq:A}\\
  \hat{B} & = \left[ -\sqrt{n} d^{\dagger} + \sqrt{1-n} e^{i(\alpha_{2} - 
    \alpha_{1})} \right] t_{1}' t_{2} \nonumber \\
  &  +e^{i \alpha_3 } \left[ -\sqrt{n} e^{i\delta} d^{\dagger} + \sqrt{1-n} 
    e^{i(\alpha_{2} - \alpha_{1})} \right] r_{1}' r_{2}' \,,
  \label{eq:B}
  \end{align}
\end{subequations}
where $\alpha_{3} \equiv  \phi_{\mathrm{u}} - 
\phi_{\mathrm{d}}-\phi_{\mathrm{AB}}$.  The current at D1
is proportional to the probability of finding there the electron,
\begin{equation}
  T = \left| t_{1}' r_{2} \right|^{2} + \left| r_{1}' t_{2}' \right|^{2} + 2 
  \left| t_{1}' r_{2} r_{1}' t_{2}' \right| \Re \left[ e^{i\alpha_{4}}   
    \left\langle e^{i \tilde{\delta}}\right\rangle \right]\,,
    \label{eq:P_D1}
\end{equation}
with $\alpha_{4} \equiv \phi_{\mathrm{u}} - \phi_{\mathrm{d}} - 
\phi_{\mathrm{AB}} + \beta_{2} - \beta_{1}$, and the phases $\beta_1$ and 
$\beta_2$ are defined by $t_1' r_2 \equiv |t_1' r_2| e^{i \beta_1}$ and $r_1' 
t_2' \equiv |r_1' t_2'| e^{i \beta_2}$. We have abbreviated $n e^{i\delta} + 
(1 - n)$ as $\langle e^{i \tilde{\delta}}\rangle$. Hereafter
$\langle \cdots \rangle$ denotes averaging with respect to the probability 
distribution function
\begin{equation}
  P ( \tilde{\delta} ) =
  \begin{cases}
    n     & \mathrm{for\; the\; phase\; to\; be\;\delta}\,;\\
    1 - n & \mathrm{for\; the\; phase\; to\; be\;0}\,.
  \end{cases}
  \label{probability_distribution_of_delta}
\end{equation}
The parameter $\delta$ is defined above, and a corresponding random variable 
is denoted by $\tilde{\delta}$. Thus, it is obtained that the part of the 
transmission phase acquired due to the interaction with the QD is a 
statistical variable whose probability distribution is determined by $n$ (cf.  
Ref.~\onlinecite{Stern_1990}).

From Eq.~(\ref{eq:P_D1}) we obtain the transmission phase through the upper 
MZI arm  and the visibility,
\begin{subequations}\label{eq:relative_phase_and_visibility}
  \begin{align}
  \arg \langle e^{i \tilde{\delta}} \rangle & = \arg \left[ n \left( e^{i 
  \delta} - 
    1 \right) + 1 \right] \,,
  \label{eq:relative_phase}\\
  v & = \frac{2}{p^{-1}+p} \left| \left \langle e^{i \tilde{\delta}} \right 
  \rangle \right| =  \frac{2 \left| 1+n (e^{i \delta} - 1)\right|}{p^{-1}+p} 
  \,,
  \label{eq:vis}
  \end{align}
\end{subequations}
with $p = |r_{1}' t_{2}' / t_{1}' r_{2}|$.

Several observations can be extracted from the results outlined above. Full 
dephasing requires $\delta = \pi$, as well as $n = 1/2$ for a particular value 
of  $\varepsilon$.  Moreover, the visibility $v(n)$ is symmetric with respect 
to $n = 1/2$, and is thus continuous at that point, even if $n(\varepsilon)$ 
is not.  As long as $\delta \ne 0$ the point $n = 1/2$ is the minimum of 
$v(n)$ (reaching $0$ for $\delta = \pi$), and $v(n)$ increases monotonically 
when $n$ moves away from that point (when $\delta = 0$ it follows that $v(n) = 
1$).  In particular, if for some reason the occupation of the QD cannot be 
$1/2$, then the visibility cannot reach zero, namely the MZI cannot be fully 
dephased.

To complete the calculation of the transmission phase and the visibility one 
needs to substitute the appropriate function for $n(\varepsilon)$. A QD 
tunnel-coupled to a Tomonaga-Luttinger liquid is analysed in 
Ref.~\onlinecite{Furusaki_2002}, where a perturbative calculation in the 
QD-lead tunneling matrix element yields $n(\varepsilon > 0) = [\Gamma /
\pi\rho_{1}(0)] \int_{0}^{\infty} \mathrm{d}\omega \rho_{g}(\omega) / (\omega
+ \varepsilon)^{2}$. Here $\rho_g (\omega)$ is the tunneling density of states 
of the Tomonaga-Luttinger liquid, and $g$ is the Luttinger parameter. If $g = 
1$, i.e. in the Fermi liquid case, $\Gamma$ takes the physical meaning of the 
level width. In our case $g \rightarrow \nu - 1$, where $\nu$ is the filling 
factor in the bulk.  For $\varepsilon < 0$  electron-hole symmetry implies $ 
n(\varepsilon) = 1 - n(-\varepsilon)$. It follows that $n(\varepsilon)$ is a 
monotonically decreasing function, and that $\mathrm{d} n / \mathrm{d} 
\varepsilon$ is a symmetric function.  The intriguing feature of 
$n(\varepsilon)$ is that the QD occupancy jumps abruptly at $\varepsilon = 0$ 
downwards if $\Gamma$ is sufficiently small (see below). This phenomenon has 
to do with the suppression of tunneling into a Tomonaga-Luttinger liquid at 
low energies, which gives rise~\cite{Furusaki_2002} to a renormalized level 
width, and under certain circumstances may result in a degenerate ground state 
of the QD+lead system.  The abrupt jump in $n(\varepsilon)$ implies that the 
value $n = 1/2$ is inaccessible as $\varepsilon$ is varied. Consequently, full 
dephasing is not possible, even when $\delta = \pi$, cf.  Eq.~(\ref{eq:vis}).  
Furthermore, according to Eq.~(\ref{eq:relative_phase}) the abrupt jump in 
$n(\varepsilon)$ is accompanied by an abrupt phase jump.

\begin{figure}[tbp]
\begin{center}
\vspace{0cm}
\includegraphics[width=0.48\textwidth]{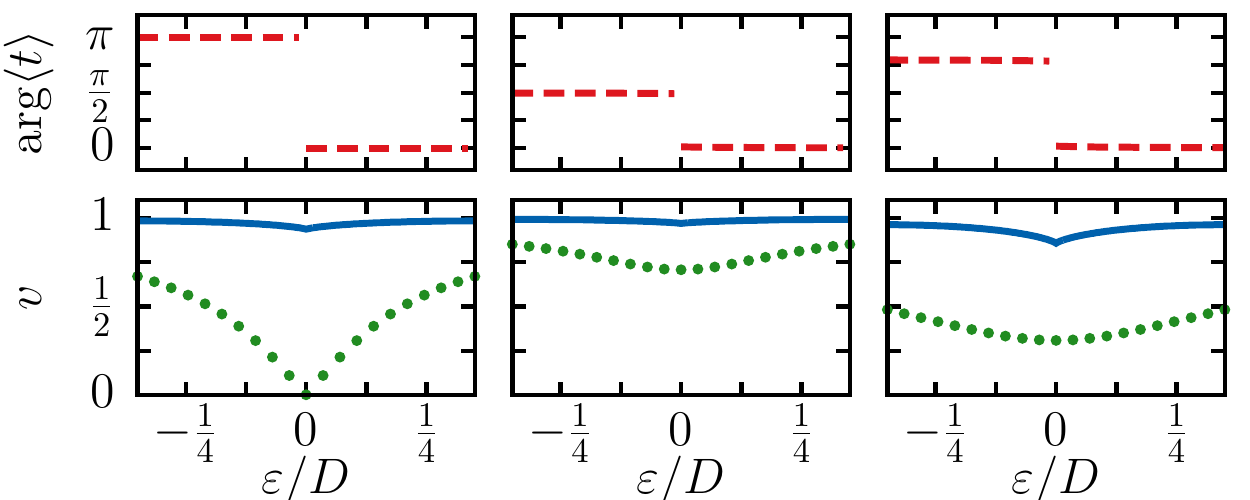}
\vspace{-0.7cm}
\caption[] {\label{fig:vis_phase_occupation}
Transmission phase through the upper MZI arm (red dashed lines, top) and its 
visibility (blue solid lines, bottom) as a function of the dimensionless 
energy parameter $\varepsilon / D$ for the fractional case.  The visibility in 
the integer case is also shown (green dotted lines, bottom). In the fractional 
case the visibility cannot vanish, but a phase lapse always occurs.  In the 
integer case the visibility can vanish, and only then a phase lapse occurs. 
Here $\Gamma / D = 0.2$, $\delta = \pi$ (left); $\Gamma / D = 0.2$, $\delta = 
\pi/2$ (center); $\Gamma / D = 0.5$, $\delta = 0.8 \pi$ (right); and $p = 1$ 
everywhere.  \vspace{-0.5cm}
}
\end{center}
\end{figure}

As an example, we focus on the case of $\nu = 4/3$, where the MZI consists of 
the chiral inner edge channels associated with the first full Landau level, 
and the QD is coupled to the   outer edge channel associated with the second, 
$1/3$-filled, Landau level.  Under these circumstances~\cite{Furusaki_2002},
$n(\varepsilon > 0) \simeq (\Gamma/2 \pi D) \left[ 1 + 2 (\varepsilon/D) \ln
(\varepsilon/D) \right]$, where $D$ is the bandwidth in the Tomonaga-Luttinger 
liquid. It follows that $n(\varepsilon)$ is discontinuous at $\varepsilon = 
0$; the downwards discontinuity is $1 - \Gamma / \pi D$. It is accompanied by 
a sharp jump in the transmission phase through the upper MZI arm whose 
magnitude is
\begin{multline}
  \arg \langle e^{i \tilde{\delta}} \rangle_{0^{+}} - \arg \langle e^{i
  \tilde{\delta}} \rangle_{0^{-}} = \arg \left[1 - \Gamma / 2 \pi D \right. \\
    + \left. (\Gamma / 2 \pi D) e^{i \delta} \right] - \arg \left[\Gamma / 2 
      \pi D + (1 - \Gamma / 2 \pi D ) e^{i \delta} \right]\, .
    \label{eq:magnitude_of_jump}
\end{multline}
Since $n \ne 1/2$, the visibility is always finite. The transmission phase and 
visibility~\footnote{See Supplemental Material for further discussion of the
transmission probability through the MZI, Eq.~(\ref{eq:P_D1}).} are depicted 
in~Fig.~\ref{fig:vis_phase_occupation}.

To emphasize the uniqueness of our results, we contrast them with a more 
trivial scenario, where the channel  tunnel-coupled to the QD supports an 
integer filling factor. Equations~(\ref{eq:relative_phase}) and (\ref{eq:vis}) 
for the transmission phase and visibility remain valid, with the proviso that 
the occupation of the QD is given by~\cite{Mahan_3ed} $n(\varepsilon) = \left[
\pi / 2 - \arctan \left( \varepsilon / \Gamma \right) \right] / \pi$.  
Substitution in Eq.~(\ref{eq:vis}) readily gives $| \langle e^{i
\tilde{\delta}} \rangle | = | 1 + (e^{i \delta} - 1) \left[ \pi/2 - \arctan 
  \left( \varepsilon / \Gamma \right) \right] / \pi |$, which is also depicted
in Fig.~\ref{fig:vis_phase_occupation}. As $n$ increases from $0$ to $1$, the 
transmission phase given by Eq.~(\ref{eq:relative_phase}) increases 
(decreases) continuously from $0$ to $\delta$ ($2\pi$ to $\delta$) if $\delta 
< \pi$  ($\delta > \pi$).  Only for $\delta = \pi$ there is an abrupt jump in 
the transmission phase, whose magnitude is $\pi$, at $n = 1/2$ (the direction 
of the jump is undefined).  The visibility is a symmetric function with 
respect to $\varepsilon = 0$, and has a minimum at that point.  If $\delta = 
\pi$, the visibility vanishes at $\varepsilon = 0$, where $n = 1/2$. Thus, in 
the integer case a phase lapse is generically absent, and full dephasing may 
occur. We note that this result can be also obtained by calculating a certain 
overlap integral associated with the QD+lead wave 
function~\cite{Rosenow_2012}.  The correspondence between this approach and 
the one we used is another manifestation of two viewpoints vis-\`a-vis 
dephasing~\cite{Stern_1990}.

What is the physical origin of the striking behaviour of the visibility in the 
fractional case as compared to the integer case? As we shall see, it is the 
degree of entanglement between the QD and the MZI. We quantify this
entanglement with the help of the \emph{concurrence}~\cite{Wootters_1998} $C$.  
In the concerned setup the concurrence is a measure of the entanglement, 
extracted from the density matrix of the system.  The latter includes all the 
information about the state of the system, and in particular on the 
entanglement between its constituents.  Larger values of the concurrence 
signify larger entanglement in the system, where $C = 0\,(1)$ implies no 
(complete) entanglement.
Using Eq.~(\ref{eq:state_in_front_of_QPC2}) and the definition of the 
concurrence~\cite{Wootters_1998}, one obtains
\begin{equation}
  C (n, \delta) = 
  4 \sqrt{n(1-n)} \left| t_{1}' \right| \sqrt{1 - \left|t_{1}' \right|^{2}} 
    \left|\sin \left( \delta/2 \right) \right| \,,
  \label{eq:concurrence}
\end{equation}
which is plotted in Fig.~\ref{fig:visbility_AAFO_concurrence} (inset). The 
concurrence (entanglement) is proportional to the quantum fluctuations in the 
occupation of the QD and to the quantum fluctuations in the  occupation of the 
MZI's arm. It has a maximum at $(n, \delta) = (1/2, \pi)$, the point of full 
dephasing.

In fact, it is possible to express the visibility as a function of the 
concurrence. This is given by
\begin{equation}
  v(C) = 2 (p + p^{-1})^{-1} [1 - C^{2}/4 \left| t_{1}' \right|^{2} (1 - 
    \left|t_{1}' \right|^{2}) ]^{\frac{1}{2}} \,,
  \label{vis_AAFO_C}
\end{equation}
and is shown in Fig.~\ref{fig:visbility_AAFO_concurrence}. This relation
has a clear physical meaning: the larger the entanglement between the MZI and 
the QD, the larger is the dephasing.  Taking the opposite point of view, we 
can say that it is possible to measure the entanglement between the two 
subsystems by measuring the visibility of the interferometer. This 
demonstrates the intimate relation between the visibility of the 
interferometer and the ability of an interfering electron to entangle to an 
electron residing on the QD.

The importance of the results reported here is twofold. From a fundamental 
point of view, we note that for a given value of $\delta$ the interaction 
between the interferometer and the QD is identical in the fractional and 
integer cases. Nevertheless, the dephasing of the interferometer by the 
QD+lead system --- and hence the entanglement between them --- is utterly 
different. This clearly underlines the importance of the internal dynamics of 
a system in the process of entangling to another system. From the experimental 
point of view, the sharp jump in the phase accompanied by high visibility can 
be used to identify unequivocally Luttinger liquid physics.  Such a 
measurement is within reach~\cite{Weisz_2012}, and the pronounced features 
should be easy to observe. We also stress that the system considered here can 
be realized in other contexts, e.g. with carbon nanotubes, where the MZI is 
replaced by a relatively open Fabry-P\'erot interferometer (so as to suppress 
multiple windings).

\begin{figure}[htb]
\begin{center}
\includegraphics[width=0.48\textwidth]{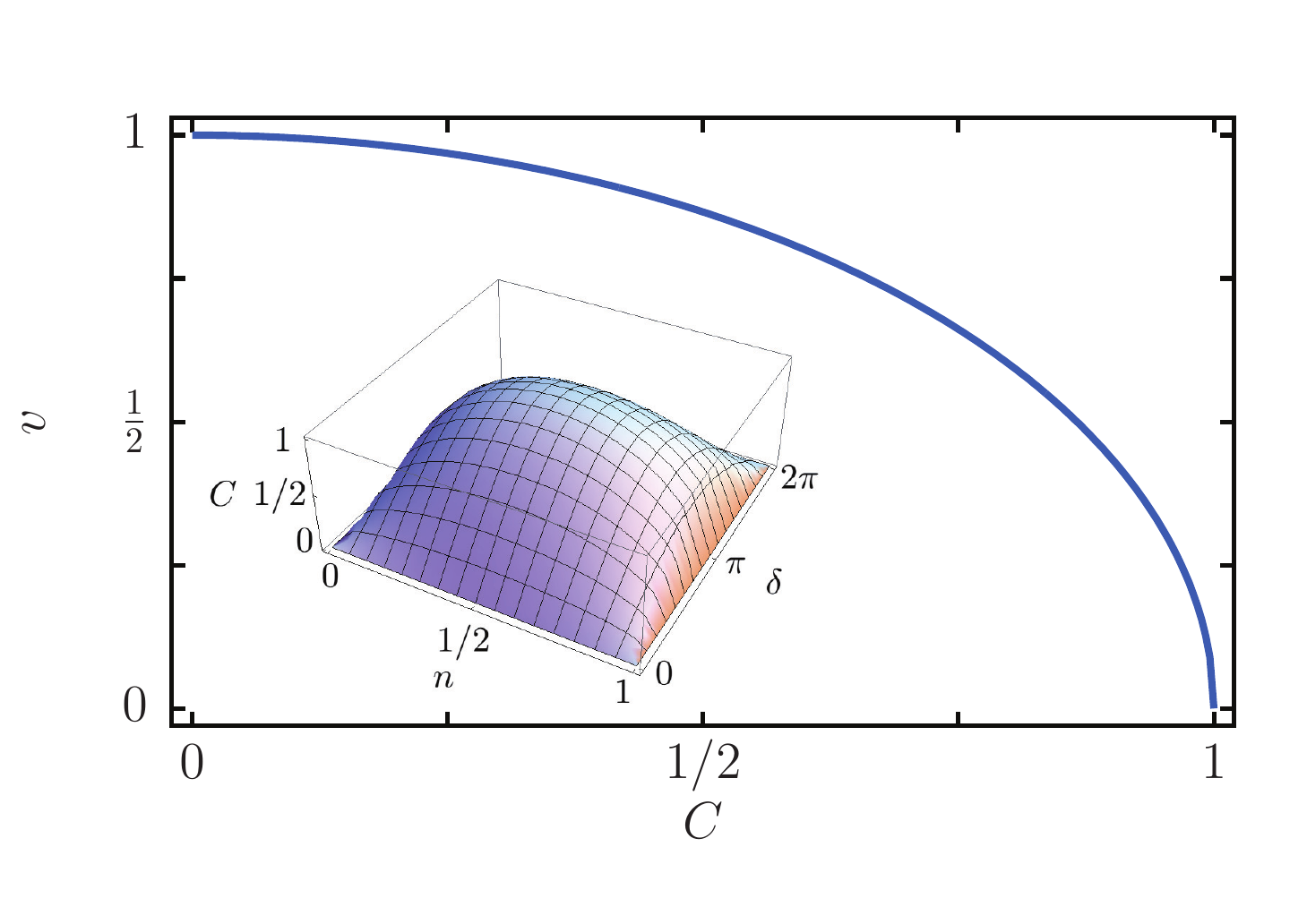}
\vspace{-1.0cm}
\caption[] {\label{fig:visbility_AAFO_concurrence}
The visibility as a function of the concurrence, $v(C)$, is a monotonically 
decreasing function (cf.  Eq.~(\ref{vis_AAFO_C})). Since, roughly, the 
concurrence is a measure of the entanglement between an electron passing 
through the MZI and an electron residing on the QD, this shows that the larger 
the entanglement between them, the larger is the dephasing of the MZI. Inset: 
The concurrence $C$ as a function of the localized state occupation $n$ and 
the screening parameter $\delta$  (cf.  Eq.~(\ref{eq:concurrence})).  The 
maximum concurrence at $(n, \delta) = (1/2, \pi)$ signifies maximum (and 
complete) entanglement between an electron interfering through the MZI and an 
electron residing on the QD. In both graphs $\left| t_1' \right| = \left| t_2' 
\right| = 1/\sqrt{2}$. }
\vspace{-0.5cm}
\end{center}
\end{figure}

\emph{To conclude}, we have analysed the interference degradation of a MZI due 
to coupling to a QD which serves as a generator of quantum telegraph noise.  
Charge fluctuations in the QD due to tunnel-coupling to a chiral edge state 
dephase the MZI. The amount of dephasing is related to the degree of 
entanglement between the MZI and the QD. We have studied the case where the QD 
is tunnel-coupled to a chiral channel supporting a fractional bulk filling and 
contrasted it with the case of an integer filling. In the former case, when 
the edge channel is modeled by a chiral Luttinger liquid, the visibility 
cannot vanish, and a phase lapse always occurs. In the integer case the 
visibility can vanish, and a phase lapse occurs only under strict symmetry 
conditions. This reduction of dephasing is shown to be intimately related to 
the fact that the system (MZI) and the detector (QD) cannot be fully 
entangled.

We are grateful to Y. Ronen, H. K. Choi, E. Weisz, Y. E. Kraus, and M.  
Heiblum for useful discussions.  Financial support by the German-Israel 
Foundation (GIF), DFG grant RO 2247/8-1, and the Israel Science Foundation is 
acknowledged.

\bibliography{ref}

\begin{thebibliography}{28}%
\makeatletter
\providecommand \@ifxundefined [1]{%
 \@ifx{#1\undefined}
}%
\providecommand \@ifnum [1]{%
 \ifnum #1\expandafter \@firstoftwo
 \else \expandafter \@secondoftwo
 \fi
}%
\providecommand \@ifx [1]{%
 \ifx #1\expandafter \@firstoftwo
 \else \expandafter \@secondoftwo
 \fi
}%
\providecommand \natexlab [1]{#1}%
\providecommand \enquote  [1]{``#1''}%
\providecommand \bibnamefont  [1]{#1}%
\providecommand \bibfnamefont [1]{#1}%
\providecommand \citenamefont [1]{#1}%
\providecommand \href@noop [0]{\@secondoftwo}%
\providecommand \href [0]{\begingroup \@sanitize@url \@href}%
\providecommand \@href[1]{\@@startlink{#1}\@@href}%
\providecommand \@@href[1]{\endgroup#1\@@endlink}%
\providecommand \@sanitize@url [0]{\catcode `\\12\catcode `\$12\catcode
  `\&12\catcode `\#12\catcode `\^12\catcode `\_12\catcode `\%12\relax}%
\providecommand \@@startlink[1]{}%
\providecommand \@@endlink[0]{}%
\providecommand \url  [0]{\begingroup\@sanitize@url \@url }%
\providecommand \@url [1]{\endgroup\@href {#1}{\urlprefix }}%
\providecommand \urlprefix  [0]{URL }%
\providecommand \Eprint [0]{\href }%
\providecommand \doibase [0]{http://dx.doi.org/}%
\providecommand \selectlanguage [0]{\@gobble}%
\providecommand \bibinfo  [0]{\@secondoftwo}%
\providecommand \bibfield  [0]{\@secondoftwo}%
\providecommand \translation [1]{[#1]}%
\providecommand \BibitemOpen [0]{}%
\providecommand \bibitemStop [0]{}%
\providecommand \bibitemNoStop [0]{.\EOS\space}%
\providecommand \EOS [0]{\spacefactor3000\relax}%
\providecommand \BibitemShut  [1]{\csname bibitem#1\endcsname}%
\let\auto@bib@innerbib\@empty
\bibitem [{\citenamefont {Ji}\ \emph {et~al.}(2003)\citenamefont {Ji},
  \citenamefont {Chung}, \citenamefont {Shprinzak}, \citenamefont {Heiblum},
  \citenamefont {Mahalu},\ and\ \citenamefont {Shtrikman}}]{Ji_2003}%
  \BibitemOpen
  \bibfield  {author} {\bibinfo {author} {\bibfnamefont {Y.}~\bibnamefont
  {Ji}}, \bibinfo {author} {\bibfnamefont {Y.}~\bibnamefont {Chung}}, \bibinfo
  {author} {\bibfnamefont {D.}~\bibnamefont {Shprinzak}}, \bibinfo {author}
  {\bibfnamefont {M.}~\bibnamefont {Heiblum}}, \bibinfo {author} {\bibfnamefont
  {D.}~\bibnamefont {Mahalu}}, \ and\ \bibinfo {author} {\bibfnamefont
  {H.}~\bibnamefont {Shtrikman}},\ }\href@noop {} {\bibfield  {journal}
  {\bibinfo  {journal} {Nature (London)}\ }\textbf {\bibinfo {volume} {422}},\
  \bibinfo {pages} {415} (\bibinfo {year} {2003})}\BibitemShut {NoStop}%
\bibitem [{\citenamefont {Neder}\ \emph {et~al.}(2006)\citenamefont {Neder},
  \citenamefont {Heiblum}, \citenamefont {Levinson}, \citenamefont {Mahalu},\
  and\ \citenamefont {Umansky}}]{Neder_2006}%
  \BibitemOpen
  \bibfield  {author} {\bibinfo {author} {\bibfnamefont {I.}~\bibnamefont
  {Neder}}, \bibinfo {author} {\bibfnamefont {M.}~\bibnamefont {Heiblum}},
  \bibinfo {author} {\bibfnamefont {Y.}~\bibnamefont {Levinson}}, \bibinfo
  {author} {\bibfnamefont {D.}~\bibnamefont {Mahalu}}, \ and\ \bibinfo {author}
  {\bibfnamefont {V.}~\bibnamefont {Umansky}},\ }\href@noop {} {\bibfield
  {journal} {\bibinfo  {journal} {Phys. Rev. Lett.}\ }\textbf {\bibinfo
  {volume} {96}},\ \bibinfo {pages} {016804} (\bibinfo {year}
  {2006})}\BibitemShut {NoStop}%
\bibitem [{\citenamefont {Roulleau}\ \emph {et~al.}(2007)\citenamefont
  {Roulleau}, \citenamefont {Portier}, \citenamefont {Glattli}, \citenamefont
  {Roche}, \citenamefont {Cavanna}, \citenamefont {Faini}, \citenamefont
  {Gennser},\ and\ \citenamefont {Mailly}}]{Roulleau_2007}%
  \BibitemOpen
  \bibfield  {author} {\bibinfo {author} {\bibfnamefont {P.}~\bibnamefont
  {Roulleau}}, \bibinfo {author} {\bibfnamefont {F.}~\bibnamefont {Portier}},
  \bibinfo {author} {\bibfnamefont {D.~C.}\ \bibnamefont {Glattli}}, \bibinfo
  {author} {\bibfnamefont {P.}~\bibnamefont {Roche}}, \bibinfo {author}
  {\bibfnamefont {A.}~\bibnamefont {Cavanna}}, \bibinfo {author} {\bibfnamefont
  {G.}~\bibnamefont {Faini}}, \bibinfo {author} {\bibfnamefont
  {U.}~\bibnamefont {Gennser}}, \ and\ \bibinfo {author} {\bibfnamefont
  {D.}~\bibnamefont {Mailly}},\ }\href@noop {} {\bibfield  {journal} {\bibinfo
  {journal} {Phys. Rev. B}\ }\textbf {\bibinfo {volume} {76}},\ \bibinfo
  {pages} {161309(R)} (\bibinfo {year} {2007})}\BibitemShut {NoStop}%
\bibitem [{\citenamefont {Litvin}\ \emph {et~al.}(2007)\citenamefont {Litvin},
  \citenamefont {Tranitz}, \citenamefont {Wegscheider},\ and\ \citenamefont
  {Strunk}}]{Litvin_2007}%
  \BibitemOpen
  \bibfield  {author} {\bibinfo {author} {\bibfnamefont {L.~V.}\ \bibnamefont
  {Litvin}}, \bibinfo {author} {\bibfnamefont {H.-P.}\ \bibnamefont {Tranitz}},
  \bibinfo {author} {\bibfnamefont {W.}~\bibnamefont {Wegscheider}}, \ and\
  \bibinfo {author} {\bibfnamefont {C.}~\bibnamefont {Strunk}},\ }\href@noop {}
  {\bibfield  {journal} {\bibinfo  {journal} {Phys. Rev. B}\ }\textbf {\bibinfo
  {volume} {75}},\ \bibinfo {pages} {033315} (\bibinfo {year}
  {2007})}\BibitemShut {NoStop}%
\bibitem [{\citenamefont {Litvin}\ \emph {et~al.}(2008)\citenamefont {Litvin},
  \citenamefont {Helzel}, \citenamefont {Tranitz}, \citenamefont
  {Wegscheider},\ and\ \citenamefont {Strunk}}]{Litvin_2008}%
  \BibitemOpen
  \bibfield  {author} {\bibinfo {author} {\bibfnamefont {L.~V.}\ \bibnamefont
  {Litvin}}, \bibinfo {author} {\bibfnamefont {A.}~\bibnamefont {Helzel}},
  \bibinfo {author} {\bibfnamefont {H.-P.}\ \bibnamefont {Tranitz}}, \bibinfo
  {author} {\bibfnamefont {W.}~\bibnamefont {Wegscheider}}, \ and\ \bibinfo
  {author} {\bibfnamefont {C.}~\bibnamefont {Strunk}},\ }\href@noop {}
  {\bibfield  {journal} {\bibinfo  {journal} {Phys. Rev. B}\ }\textbf {\bibinfo
  {volume} {78}},\ \bibinfo {pages} {075303} (\bibinfo {year}
  {2008})}\BibitemShut {NoStop}%
\bibitem [{\citenamefont {Roulleau}\ \emph {et~al.}(2008)\citenamefont
  {Roulleau}, \citenamefont {Portier}, \citenamefont {Roche}, \citenamefont
  {Cavanna}, \citenamefont {Faini}, \citenamefont {Gennser},\ and\
  \citenamefont {Mailly}}]{Roulleau_2008}%
  \BibitemOpen
  \bibfield  {author} {\bibinfo {author} {\bibfnamefont {P.}~\bibnamefont
  {Roulleau}}, \bibinfo {author} {\bibfnamefont {F.}~\bibnamefont {Portier}},
  \bibinfo {author} {\bibfnamefont {P.}~\bibnamefont {Roche}}, \bibinfo
  {author} {\bibfnamefont {A.}~\bibnamefont {Cavanna}}, \bibinfo {author}
  {\bibfnamefont {G.}~\bibnamefont {Faini}}, \bibinfo {author} {\bibfnamefont
  {U.}~\bibnamefont {Gennser}}, \ and\ \bibinfo {author} {\bibfnamefont
  {D.}~\bibnamefont {Mailly}},\ }\href@noop {} {\bibfield  {journal} {\bibinfo
  {journal} {Phys. Rev. Lett.}\ }\textbf {\bibinfo {volume} {100}},\ \bibinfo
  {pages} {126802} (\bibinfo {year} {2008})}\BibitemShut {NoStop}%
\bibitem [{\citenamefont {Bieri}\ \emph {et~al.}(2009)\citenamefont {Bieri},
  \citenamefont {Weiss}, \citenamefont {Goktas}, \citenamefont {Hauser},
  \citenamefont {Schonenberger},\ and\ \citenamefont
  {Oberholzer}}]{Bieri_2009}%
  \BibitemOpen
  \bibfield  {author} {\bibinfo {author} {\bibfnamefont {E.}~\bibnamefont
  {Bieri}}, \bibinfo {author} {\bibfnamefont {M.}~\bibnamefont {Weiss}},
  \bibinfo {author} {\bibfnamefont {O.}~\bibnamefont {Goktas}}, \bibinfo
  {author} {\bibfnamefont {M.}~\bibnamefont {Hauser}}, \bibinfo {author}
  {\bibfnamefont {C.}~\bibnamefont {Schonenberger}}, \ and\ \bibinfo {author}
  {\bibfnamefont {S.}~\bibnamefont {Oberholzer}},\ }\href@noop {} {\bibfield
  {journal} {\bibinfo  {journal} {Phys. Rev. B}\ }\textbf {\bibinfo {volume}
  {79}},\ \bibinfo {pages} {245324} (\bibinfo {year} {2009})}\BibitemShut
  {NoStop}%
\bibitem [{\citenamefont {Roulleau}\ \emph {et~al.}(2009)\citenamefont
  {Roulleau}, \citenamefont {Portier}, \citenamefont {Roche}, \citenamefont
  {Cavanna}, \citenamefont {Faini}, \citenamefont {Gennser},\ and\
  \citenamefont {Mailly}}]{Roulleau_2009}%
  \BibitemOpen
  \bibfield  {author} {\bibinfo {author} {\bibfnamefont {P.}~\bibnamefont
  {Roulleau}}, \bibinfo {author} {\bibfnamefont {F.}~\bibnamefont {Portier}},
  \bibinfo {author} {\bibfnamefont {P.}~\bibnamefont {Roche}}, \bibinfo
  {author} {\bibfnamefont {A.}~\bibnamefont {Cavanna}}, \bibinfo {author}
  {\bibfnamefont {G.}~\bibnamefont {Faini}}, \bibinfo {author} {\bibfnamefont
  {U.}~\bibnamefont {Gennser}}, \ and\ \bibinfo {author} {\bibfnamefont
  {D.}~\bibnamefont {Mailly}},\ }\href@noop {} {\bibfield  {journal} {\bibinfo
  {journal} {Phys. Rev. Lett.}\ }\textbf {\bibinfo {volume} {102}},\ \bibinfo
  {pages} {236802} (\bibinfo {year} {2009})}\BibitemShut {NoStop}%
\bibitem [{\citenamefont {Camino}\ \emph {et~al.}(2007)\citenamefont {Camino},
  \citenamefont {Zhou},\ and\ \citenamefont {Goldman}}]{Camino_2007}%
  \BibitemOpen
  \bibfield  {author} {\bibinfo {author} {\bibfnamefont {F.~E.}\ \bibnamefont
  {Camino}}, \bibinfo {author} {\bibfnamefont {W.}~\bibnamefont {Zhou}}, \ and\
  \bibinfo {author} {\bibfnamefont {V.~J.}\ \bibnamefont {Goldman}},\ }\href
  {\doibase 10.1103/PhysRevB.76.155305} {\bibfield  {journal} {\bibinfo
  {journal} {Phys. Rev. B}\ }\textbf {\bibinfo {volume} {76}},\ \bibinfo
  {pages} {155305} (\bibinfo {year} {2007})}\BibitemShut {NoStop}%
\bibitem [{\citenamefont {Zhang}\ \emph {et~al.}(2009)\citenamefont {Zhang},
  \citenamefont {McClure}, \citenamefont {Levenson-Falk}, \citenamefont
  {Marcus}, \citenamefont {Pfeiffer},\ and\ \citenamefont {West}}]{Zhang_2009}%
  \BibitemOpen
  \bibfield  {author} {\bibinfo {author} {\bibfnamefont {Y.}~\bibnamefont
  {Zhang}}, \bibinfo {author} {\bibfnamefont {D.~T.}\ \bibnamefont {McClure}},
  \bibinfo {author} {\bibfnamefont {E.~M.}\ \bibnamefont {Levenson-Falk}},
  \bibinfo {author} {\bibfnamefont {C.~M.}\ \bibnamefont {Marcus}}, \bibinfo
  {author} {\bibfnamefont {L.~N.}\ \bibnamefont {Pfeiffer}}, \ and\ \bibinfo
  {author} {\bibfnamefont {K.~W.}\ \bibnamefont {West}},\ }\href {\doibase
  10.1103/PhysRevB.79.241304} {\bibfield  {journal} {\bibinfo  {journal} {Phys.
  Rev. B}\ }\textbf {\bibinfo {volume} {79}},\ \bibinfo {pages} {241304}
  (\bibinfo {year} {2009})}\BibitemShut {NoStop}%
\bibitem [{\citenamefont {Ofek}\ \emph {et~al.}(2010)\citenamefont {Ofek},
  \citenamefont {Bid}, \citenamefont {Heiblum}, \citenamefont {Stern},
  \citenamefont {Umansky},\ and\ \citenamefont {Mahalu}}]{Ofek_2010}%
  \BibitemOpen
  \bibfield  {author} {\bibinfo {author} {\bibfnamefont {N.}~\bibnamefont
  {Ofek}}, \bibinfo {author} {\bibfnamefont {A.}~\bibnamefont {Bid}}, \bibinfo
  {author} {\bibfnamefont {M.}~\bibnamefont {Heiblum}}, \bibinfo {author}
  {\bibfnamefont {A.}~\bibnamefont {Stern}}, \bibinfo {author} {\bibfnamefont
  {V.}~\bibnamefont {Umansky}}, \ and\ \bibinfo {author} {\bibfnamefont
  {D.}~\bibnamefont {Mahalu}},\ }\href@noop {} {\bibfield  {journal} {\bibinfo
  {journal} {Proceedings of the National Academy of Sciences}\ }\textbf
  {\bibinfo {volume} {107}},\ \bibinfo {pages} {5276} (\bibinfo {year}
  {2010})}\BibitemShut {NoStop}%
\bibitem [{\citenamefont {McClure}\ \emph {et~al.}(2012)\citenamefont
  {McClure}, \citenamefont {Chang}, \citenamefont {Marcus}, \citenamefont
  {Pfeiffer},\ and\ \citenamefont {West}}]{McClure_2012}%
  \BibitemOpen
  \bibfield  {author} {\bibinfo {author} {\bibfnamefont {D.~T.}\ \bibnamefont
  {McClure}}, \bibinfo {author} {\bibfnamefont {W.}~\bibnamefont {Chang}},
  \bibinfo {author} {\bibfnamefont {C.~M.}\ \bibnamefont {Marcus}}, \bibinfo
  {author} {\bibfnamefont {L.~N.}\ \bibnamefont {Pfeiffer}}, \ and\ \bibinfo
  {author} {\bibfnamefont {K.~W.}\ \bibnamefont {West}},\ }\href {\doibase
  10.1103/PhysRevLett.108.256804} {\bibfield  {journal} {\bibinfo  {journal}
  {Phys. Rev. Lett.}\ }\textbf {\bibinfo {volume} {108}},\ \bibinfo {pages}
  {256804} (\bibinfo {year} {2012})}\BibitemShut {NoStop}%
\bibitem [{\citenamefont {Neder}\ \emph
  {et~al.}(2007{\natexlab{a}})\citenamefont {Neder}, \citenamefont {Heiblum},
  \citenamefont {Mahalu},\ and\ \citenamefont {Umansky}}]{Neder_2007_PRL}%
  \BibitemOpen
  \bibfield  {author} {\bibinfo {author} {\bibfnamefont {I.}~\bibnamefont
  {Neder}}, \bibinfo {author} {\bibfnamefont {M.}~\bibnamefont {Heiblum}},
  \bibinfo {author} {\bibfnamefont {D.}~\bibnamefont {Mahalu}}, \ and\ \bibinfo
  {author} {\bibfnamefont {V.}~\bibnamefont {Umansky}},\ }\href {\doibase
  10.1103/PhysRevLett.98.036803} {\bibfield  {journal} {\bibinfo  {journal}
  {Phys. Rev. Lett.}\ }\textbf {\bibinfo {volume} {98}},\ \bibinfo {pages}
  {036803} (\bibinfo {year} {2007}{\natexlab{a}})}\BibitemShut {NoStop}%
\bibitem [{\citenamefont {Neder}\ \emph
  {et~al.}(2007{\natexlab{b}})\citenamefont {Neder}, \citenamefont {Marquardt},
  \citenamefont {Heiblum}, \citenamefont {Mahalu},\ and\ \citenamefont
  {Umansky}}]{Neder_2007_NaturePhysics}%
  \BibitemOpen
  \bibfield  {author} {\bibinfo {author} {\bibfnamefont {I.}~\bibnamefont
  {Neder}}, \bibinfo {author} {\bibfnamefont {F.}~\bibnamefont {Marquardt}},
  \bibinfo {author} {\bibfnamefont {M.}~\bibnamefont {Heiblum}}, \bibinfo
  {author} {\bibfnamefont {D.}~\bibnamefont {Mahalu}}, \ and\ \bibinfo {author}
  {\bibfnamefont {V.}~\bibnamefont {Umansky}},\ }\href@noop {} {\bibfield
  {journal} {\bibinfo  {journal} {Nat. Phys.}\ }\textbf {\bibinfo {volume}
  {3}},\ \bibinfo {pages} {534} (\bibinfo {year}
  {2007}{\natexlab{b}})}\BibitemShut {NoStop}%
\bibitem [{\citenamefont {Levinson}(1997)}]{Levinson_1997}%
  \BibitemOpen
  \bibfield  {author} {\bibinfo {author} {\bibfnamefont {Y.}~\bibnamefont
  {Levinson}},\ }\href@noop {} {\bibfield  {journal} {\bibinfo  {journal}
  {Europhys. Lett.}\ }\textbf {\bibinfo {volume} {39}},\ \bibinfo {pages} {299}
  (\bibinfo {year} {1997})}\BibitemShut {NoStop}%
\bibitem [{\citenamefont {Aleiner}\ \emph {et~al.}(1997)\citenamefont
  {Aleiner}, \citenamefont {Wingreen},\ and\ \citenamefont
  {Meir}}]{Aleiner_1997}%
  \BibitemOpen
  \bibfield  {author} {\bibinfo {author} {\bibfnamefont {I.~L.}\ \bibnamefont
  {Aleiner}}, \bibinfo {author} {\bibfnamefont {N.~S.}\ \bibnamefont
  {Wingreen}}, \ and\ \bibinfo {author} {\bibfnamefont {Y.}~\bibnamefont
  {Meir}},\ }\href@noop {} {\bibfield  {journal} {\bibinfo  {journal} {Phys.
  Rev. Lett.}\ }\textbf {\bibinfo {volume} {79}},\ \bibinfo {pages} {3740}
  (\bibinfo {year} {1997})}\BibitemShut {NoStop}%
\bibitem [{\citenamefont {Gurvitz}(1997)}]{Gurvitz_1997}%
  \BibitemOpen
  \bibfield  {author} {\bibinfo {author} {\bibfnamefont {S.~A.}\ \bibnamefont
  {Gurvitz}},\ }\href {\doibase 10.1103/PhysRevB.56.15215} {\bibfield
  {journal} {\bibinfo  {journal} {Phys. Rev. B}\ }\textbf {\bibinfo {volume}
  {56}},\ \bibinfo {pages} {15215} (\bibinfo {year} {1997})}\BibitemShut
  {NoStop}%
\bibitem [{\citenamefont {Rosenow}\ and\ \citenamefont
  {Gefen}(2012)}]{Rosenow_2012}%
  \BibitemOpen
  \bibfield  {author} {\bibinfo {author} {\bibfnamefont {B.}~\bibnamefont
  {Rosenow}}\ and\ \bibinfo {author} {\bibfnamefont {Y.}~\bibnamefont
  {Gefen}},\ }\href {\doibase 10.1103/PhysRevLett.108.256805} {\bibfield
  {journal} {\bibinfo  {journal} {Phys. Rev. Lett.}\ }\textbf {\bibinfo
  {volume} {108}},\ \bibinfo {pages} {256805} (\bibinfo {year}
  {2012})}\BibitemShut {NoStop}%
\bibitem [{\citenamefont {Grishin}\ \emph {et~al.}(2005)\citenamefont
  {Grishin}, \citenamefont {Yurkevich},\ and\ \citenamefont
  {Lerner}}]{Grishin_2005}%
  \BibitemOpen
  \bibfield  {author} {\bibinfo {author} {\bibfnamefont {A.}~\bibnamefont
  {Grishin}}, \bibinfo {author} {\bibfnamefont {I.~V.}\ \bibnamefont
  {Yurkevich}}, \ and\ \bibinfo {author} {\bibfnamefont {I.~V.}\ \bibnamefont
  {Lerner}},\ }\href {\doibase 10.1103/PhysRevB.72.060509} {\bibfield
  {journal} {\bibinfo  {journal} {Phys. Rev. B}\ }\textbf {\bibinfo {volume}
  {72}},\ \bibinfo {pages} {060509} (\bibinfo {year} {2005})}\BibitemShut
  {NoStop}%
\bibitem [{\citenamefont {Yurkevich}\ \emph {et~al.}(2010)\citenamefont
  {Yurkevich}, \citenamefont {Baldwin}, \citenamefont {Lerner},\ and\
  \citenamefont {Altshuler}}]{Yurkevich_2010}%
  \BibitemOpen
  \bibfield  {author} {\bibinfo {author} {\bibfnamefont {I.~V.}\ \bibnamefont
  {Yurkevich}}, \bibinfo {author} {\bibfnamefont {J.}~\bibnamefont {Baldwin}},
  \bibinfo {author} {\bibfnamefont {I.~V.}\ \bibnamefont {Lerner}}, \ and\
  \bibinfo {author} {\bibfnamefont {B.~L.}\ \bibnamefont {Altshuler}},\ }\href
  {\doibase 10.1103/PhysRevB.81.121305} {\bibfield  {journal} {\bibinfo
  {journal} {Phys. Rev. B}\ }\textbf {\bibinfo {volume} {81}},\ \bibinfo
  {pages} {121305} (\bibinfo {year} {2010})}\BibitemShut {NoStop}%
\bibitem [{\citenamefont {Neder}\ and\ \citenamefont
  {Marquardt}(2007)}]{Neder_2007_NewJournalOfPhysics}%
  \BibitemOpen
  \bibfield  {author} {\bibinfo {author} {\bibfnamefont {I.}~\bibnamefont
  {Neder}}\ and\ \bibinfo {author} {\bibfnamefont {F.}~\bibnamefont
  {Marquardt}},\ }\href@noop {} {\bibfield  {journal} {\bibinfo  {journal} {New
  J. Phys.}\ }\textbf {\bibinfo {volume} {9}},\ \bibinfo {pages} {112}
  (\bibinfo {year} {2007})}\BibitemShut {NoStop}%
\bibitem [{\citenamefont {Abel}\ and\ \citenamefont
  {Marquardt}(2008)}]{Abel_2008}%
  \BibitemOpen
  \bibfield  {author} {\bibinfo {author} {\bibfnamefont {B.}~\bibnamefont
  {Abel}}\ and\ \bibinfo {author} {\bibfnamefont {F.}~\bibnamefont
  {Marquardt}},\ }\href {\doibase 10.1103/PhysRevB.78.201302} {\bibfield
  {journal} {\bibinfo  {journal} {Phys. Rev. B}\ }\textbf {\bibinfo {volume}
  {78}},\ \bibinfo {pages} {201302} (\bibinfo {year} {2008})}\BibitemShut
  {NoStop}%
\bibitem [{\citenamefont {Weisz}\ \emph {et~al.}(2012)\citenamefont {Weisz},
  \citenamefont {Choi}, \citenamefont {Heiblum}, \citenamefont {Gefen},
  \citenamefont {Umansky},\ and\ \citenamefont {Mahalu}}]{Weisz_2012}%
  \BibitemOpen
  \bibfield  {author} {\bibinfo {author} {\bibfnamefont {E.}~\bibnamefont
  {Weisz}}, \bibinfo {author} {\bibfnamefont {H.~K.}\ \bibnamefont {Choi}},
  \bibinfo {author} {\bibfnamefont {M.}~\bibnamefont {Heiblum}}, \bibinfo
  {author} {\bibfnamefont {Y.}~\bibnamefont {Gefen}}, \bibinfo {author}
  {\bibfnamefont {V.}~\bibnamefont {Umansky}}, \ and\ \bibinfo {author}
  {\bibfnamefont {D.}~\bibnamefont {Mahalu}},\ }\href {\doibase
  10.1103/PhysRevLett.109.250401} {\bibfield  {journal} {\bibinfo  {journal}
  {Phys. Rev. Lett.}\ }\textbf {\bibinfo {volume} {109}},\ \bibinfo {pages}
  {250401} (\bibinfo {year} {2012})}\BibitemShut {NoStop}%
\bibitem [{\citenamefont {Stern}\ \emph {et~al.}(1990)\citenamefont {Stern},
  \citenamefont {Aharonov},\ and\ \citenamefont {Imry}}]{Stern_1990}%
  \BibitemOpen
  \bibfield  {author} {\bibinfo {author} {\bibfnamefont {A.}~\bibnamefont
  {Stern}}, \bibinfo {author} {\bibfnamefont {Y.}~\bibnamefont {Aharonov}}, \
  and\ \bibinfo {author} {\bibfnamefont {Y.}~\bibnamefont {Imry}},\ }\href
  {\doibase 10.1103/PhysRevA.41.3436} {\bibfield  {journal} {\bibinfo
  {journal} {Phys. Rev. A}\ }\textbf {\bibinfo {volume} {41}},\ \bibinfo
  {pages} {3436} (\bibinfo {year} {1990})}\BibitemShut {NoStop}%
\bibitem [{\citenamefont {Furusaki}\ and\ \citenamefont
  {Matveev}(2002)}]{Furusaki_2002}%
  \BibitemOpen
  \bibfield  {author} {\bibinfo {author} {\bibfnamefont {A.}~\bibnamefont
  {Furusaki}}\ and\ \bibinfo {author} {\bibfnamefont {K.~A.}\ \bibnamefont
  {Matveev}},\ }\href {\doibase 10.1103/PhysRevLett.88.226404} {\bibfield
  {journal} {\bibinfo  {journal} {Phys. Rev. Lett.}\ }\textbf {\bibinfo
  {volume} {88}},\ \bibinfo {pages} {226404} (\bibinfo {year}
  {2002})}\BibitemShut {NoStop}%
\bibitem [{Note1()}]{Note1}%
  \BibitemOpen
  \bibinfo {note} {See Supplemental Material for further discussion of the
  transmission probability through the MZI, Eq.~(\ref {eq:P_D1}).}\BibitemShut
  {Stop}%
\bibitem [{\citenamefont {Mahan}(2000)}]{Mahan_3ed}%
  \BibitemOpen
  \bibfield  {author} {\bibinfo {author} {\bibfnamefont {G.~D.}\ \bibnamefont
  {Mahan}},\ }\href@noop {} {\emph {\bibinfo {title} {Many-Particle
  Physics}}},\ \bibinfo {edition} {3rd}\ ed.\ (\bibinfo  {publisher} {Kluwer
  Academic / Plenum Publishers},\ \bibinfo {address} {New York},\ \bibinfo
  {year} {2000})\ Chap.\ \bibinfo {chapter} {4.2}\BibitemShut {NoStop}%
\bibitem [{\citenamefont {Wootters}(1998)}]{Wootters_1998}%
  \BibitemOpen
  \bibfield  {author} {\bibinfo {author} {\bibfnamefont {W.~K.}\ \bibnamefont
  {Wootters}},\ }\href {\doibase 10.1103/PhysRevLett.80.2245} {\bibfield
  {journal} {\bibinfo  {journal} {Phys. Rev. Lett.}\ }\textbf {\bibinfo
  {volume} {80}},\ \bibinfo {pages} {2245} (\bibinfo {year}
  {1998})}\BibitemShut {NoStop}%
\end{thebibliography}%

\onecolumngrid

\pagebreak

\appendix

\section*{Supplemental Material}

\twocolumngrid

\setcounter{enumi}{1}
\setcounter{equation}{0}
\setcounter{figure}{0}
\renewcommand{\theequation}{\Roman{enumi}.\arabic{equation}}
\renewcommand{\thefigure}{\Roman{enumi}.\arabic{figure}}

\subsection*{The transmission probability through the Mach-Zehnder 
interferometer}

In the main text we discuss the visibility of the interference signal through 
the Mach-Zehnder interferometer (MZI) and the transmission phase through its 
``upper'' arm.  Both these quantities can be extracted from the transmission 
probability given by Eq.~(4) in the main text. These are the physically 
relevant quantities, since they are independent of the specific value of the 
Aharonov-Bohm flux $\Phi$, which serves as a means of measuring the 
transmission phase.  Here, we provide further information on Eq.~(4), which 
may facilitate an experimental investigation of the setup.

Eq.~(4) presents the transmission probability through the MZI, which is 
directly measured experimentally. It depends on the scattering matrices of the 
junctions of the MZI, as well as on the phase $\alpha_4$ (see text after 
Eq.~(4)). However, the physics referred to in the main text is essentially 
given by $\langle e^{i \tilde{\delta}} \rangle$. Thus, in what follows we take 
$|t'_1 r_2| = |r'_1 t'_2| = 1/2$ and $\alpha_4 = 0$, which yield
\begin{equation}
  T = \frac{1}{2} + \frac{1}{2} | \langle e^{i \tilde{\delta}} \rangle | \cos 
  \left( \arg \langle e^{i \tilde{\delta}} \rangle \right) \, .
  \label{eq:trans}
\end{equation}
This is depicted in Fig.~\ref{fig:probability} for the three cases shown in 
Fig.~(2), both for the fractional and integer regimes.

As can be seen from Fig.~\ref{fig:probability}, and as can be also verified by 
Eq.~(4), an abrupt jump in the transmission phase leads to an abrupt jump in 
the transmission probability through the MZI. Thus for the integer case the 
transmission probability is continuous (when $\delta = \pi$ the transmission 
phase jumps abruptly, but at the same point the visibility vanishes, thus 
rendering the jump invisible). In contrast, in the fractional case the 
transmission probability always jumps abruptly.

\vspace{0.5cm}
\begin{center}
\begin{figurehere}
\includegraphics[width=0.45\textwidth]{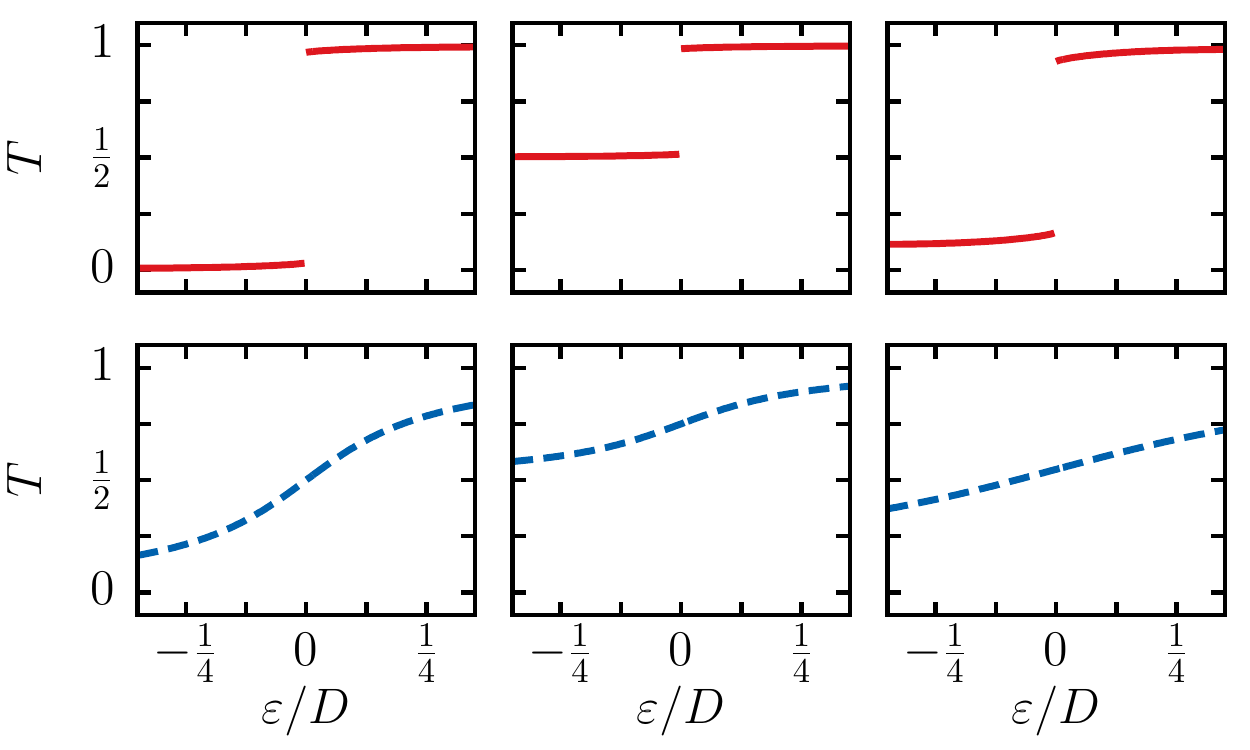}
\caption[] {\label{fig:probability}
The transmission probability through the Mach-Zehnder interferometer as a 
function of the dimensionless energy parameter $\varepsilon / D$ for the 
fractional case (upper panel) and integer case (lower panel), see 
Eq.~(\ref{eq:trans}). The abrupt jump is the hallmark of the fractional case.  
Here $\Gamma / D = 0.2$, $\delta = \pi$ (left); $\Gamma / D = 0.2$, $\delta = 
\pi/2$ (center); $\Gamma / D = 0.5$, $\delta = 0.8 \pi$ (right); $|t'_1 r_2| = 
|r'_1 t'_2| = 1/2$ and $\alpha_4 = 0$ everywhere.
} 
\end{figurehere}
\end{center}

We have focused on the transmission probability from S2 to D1. We note that 
other transmission probabilities may be calculated similarly, i.e. the 
transmission probabilities from S1 to D1, from S1 to D2, and from S2 to D2.  
However, the same physics discussed here for the pair S2-D1 holds for other 
pairs. So considering, for example, the transmission probability from S2 to 
D2, does not add any qualitatively new information.

\end{document}